\documentclass[aip,reprint]{revtex4-2}
\usepackage{comment}
\usepackage{pgfplots}
\usepgfplotslibrary{statistics}
\usepackage{amsmath}
\usepackage{amssymb}

\usepackage{graphicx}
\usepackage{dcolumn}
\usepackage{bm}
\usepackage[mathlines]{lineno}
\usepackage{nicefrac}
\usepackage[utf8]{inputenc}
\usepackage[T1]{fontenc}
\usepackage{mathptmx}
\usepackage{etoolbox}
\usepackage{csquotes}

\usepackage[sort&compress]{natbib}
\setcitestyle{numbers,square,comma,sort&compress}

\makeatletter
\def\@email#1#2{%
 \endgroup
 \patchcmd{\titleblock@produce}
  {\frontmatter@RRAPformat}
  {\frontmatter@RRAPformat{\produce@RRAP{*#1\href{mailto:#2}{#2}}}\frontmatter@RRAPformat}
  {}{}
}%
\makeatother

\usepackage{siunitx}
\DeclareSIUnit\rpm{rpm}
\DeclareSIUnit{\wtpercent}{wt\%}
\DeclareSIUnit{\px}{px}
\usepackage{microtype}

\usepackage{xfp}

\usepackage{subfigure}

\usepackage{latexsym}      
\usepackage{hyperref}
\hypersetup{
 colorlinks=false,
 citecolor=false,
 linkcolor=false,
 urlcolor=false,
hidelinks,
 pdfpagemode=UseNone,
 pdfstartview=FitH}
    
\usepackage[eulergreek]{sansmath}
\usepackage{upgreek}

\usepackage{filecontents}

\usepackage{booktabs}

\usepackage{dcolumn}
\newcolumntype{d}[1]{D{.}{.}{#1}}

\let\vec\boldsymbol
\let\bs\boldsymbol

\usepackage{xcolor}
\definecolor{c1}{rgb}{0.7068574918274737, 0.11027871818526241, 0.2747061222663145}%
\definecolor{c2}{rgb}{0.6780437401750237, 0.1857033496010309, 0.10475664313058389}%
\definecolor{c3}{rgb}{0.5819819292481591, 0.28135510723834917, 0.10365807489974799}%
\definecolor{c4}{rgb}{0.5234892407222805, 0.3183040932830017, 0.10308831855626377}%
\definecolor{c5}{rgb}{0.4801916866157751, 0.3399605081271155, 0.10271364194308724}%
\definecolor{c6}{rgb}{0.44359832760663015, 0.3553988979685888, 0.10242748858902176}%
\definecolor{c7}{rgb}{0.40913672397205286, 0.36794412723413256, 0.10218299803254591}%
\definecolor{c8}{rgb}{0.37320211354661986, 0.37925108836394167, 0.10195327596798023}%
\definecolor{c9}{rgb}{0.33135856975649947, 0.3904277497779026, 0.10171733001337946}%
\definecolor{c10}{rgb}{0.2751398809612443, 0.40252912490464393, 0.10145175401497963}%
\definecolor{c11}{rgb}{0.17705545280426335, 0.4169982861326329, 0.10112016988851782}%
\definecolor{c12}{rgb}{0.10370113411519366, 0.41987331701008007, 0.20784080256789766}%
\definecolor{c13}{rgb}{0.10672620229256541, 0.415683426104876, 0.2824387977867041}%
\definecolor{c14}{rgb}{0.10896052083074859, 0.412482459879469, 0.3262929409150867}%
\definecolor{c15}{rgb}{0.11082049183583692, 0.4097465781098524, 0.3585658834499722}%
\definecolor{c16}{rgb}{0.1125289398766243, 0.40717495121262287, 0.38576967807315987}%
\definecolor{c17}{rgb}{0.11424591411818866, 0.4045324469356048, 0.41127417284765605}%
\definecolor{c18}{rgb}{0.11613475317261168, 0.4015563987219263, 0.4376207984793854}%
\definecolor{c19}{rgb}{0.11843118069754927, 0.3978377246048391, 0.46769849466738594}%
\definecolor{c20}{rgb}{0.1215889511379443, 0.3925371130865723, 0.5062751486798138}%
\definecolor{c21}{rgb}{0.12675667510032929, 0.38336665224260497, 0.564172135389151}%
\definecolor{c22}{rgb}{0.13823697094516182, 0.36050804998003744, 0.6775165660716256}%
\definecolor{c23}{rgb}{0.2929229731919379, 0.2517707788576924, 0.9106622939003164}%
\definecolor{c24}{rgb}{0.5044517963791875, 0.15823641315504755, 0.8374462940274711}%
\definecolor{c25}{rgb}{0.5743505600217842, 0.14563681653078617, 0.7277475942695497}%
\definecolor{c26}{rgb}{0.6114384687170589, 0.13753415571759905, 0.6502354669913063}%
\definecolor{c27}{rgb}{0.6363457404298491, 0.13141740987926892, 0.586320806947322}%
\definecolor{c28}{rgb}{0.6557132733878439, 0.12622710616005786, 0.5268534496956088}%
\definecolor{c29}{rgb}{0.6725336221941806, 0.12137095721249477, 0.4649001891758376}%
\definecolor{c30}{rgb}{0.688593421429631, 0.11639483989072555, 0.39157406584043325}%
\definecolor{brickred}{rgb}{0.8, 0.25, 0.33}%
\definecolor{darkorange}{rgb}{1.0, 0.55, 0.0}%
\definecolor{persiangreen}{rgb}{0.0, 0.65, 0.58}%
\definecolor{persianindigo}{rgb}{0.2, 0.07, 0.48}%
\definecolor{cadet}{rgb}{0.33, 0.41, 0.47}%
\definecolor{turquoisegreen}{rgb}{0.63, 0.84, 0.71}%
\definecolor{sandybrown}{rgb}{0.96, 0.64, 0.38}%
\definecolor{blueblue}{rgb}{0.0, 0.2, 0.6}%
\definecolor{ballblue}{rgb}{0.13, 0.67, 0.8}%
\definecolor{greengreen}{rgb}{0.0, 0.5, 0.0}%
\definecolor{forestgreen}{rgb}{0.0, 0.42, 0.24}
\definecolor{amber}{rgb}{1.0, 0.75, 0.0}%
\definecolor{goldenrod}{rgb}{0.85, 0.65, 0.13}%
\definecolor{gold}{rgb}{0.86, 0.71, 0.23}%
\definecolor{tiffanyblue}{rgb}{0.04, 0.73, 0.71}%
\definecolor{bittersweet}{rgb}{1.0, 0.44, 0.37}%
\definecolor{darkgreen}{rgb}{0.12, 0.3, 0.17}%
\definecolor{carmine}{rgb}{0.92, 0.3, 0.26}%

\definecolor{purpleheart}{rgb}{0.41, 0.21, 0.61}%
\definecolor{richblack}{rgb}{0.0, 0.25, 0.25}
\definecolor{royalazure}{rgb}{0.0, 0.22, 0.66}
\definecolor{sacramentostategreen}{rgb}{0.0, 0.34, 0.25}

\definecolor{tyrianpurple}{rgb}{0.4, 0.01, 0.24}
\definecolor{redviolet}{rgb}{0.78, 0.08, 0.52}
\definecolor{vividburgundy}{rgb}{0.62, 0.11, 0.21}
\definecolor{amaranth}{rgb}{0.9, 0.17, 0.31}

\definecolor{ca}{RGB}{127,59,8}%
\definecolor{cb}{RGB}{179,88,6}%
\definecolor{cc}{RGB}{224,130,20}%
\definecolor{cd}{RGB}{253,184,99}%
\definecolor{ce}{RGB}{254,224,182}
\definecolor{cf}{RGB}{216,218,235}%
\definecolor{cg}{RGB}{178,171,210}%
\definecolor{ch}{RGB}{128,115,172}%
\definecolor{ci}{RGB}{84,39,136}%
\definecolor{cj}{RGB}{45,0,75}%

\definecolor{brickred}{rgb}{0.8, 0.25, 0.33}
\usepackage[colorinlistoftodos,textcolor=brickred,backgroundcolor=white,bordercolor=brickred]{todonotes}
\let\TODO\todo

\usepackage{tikz}
\usetikzlibrary{math}
\usetikzlibrary{positioning}
\usetikzlibrary{shapes.geometric, arrows}
\usetikzlibrary{arrows.meta}
\usetikzlibrary{positioning}
\usetikzlibrary{shapes,arrows}
\usetikzlibrary{intersections}
\usetikzlibrary{patterns}
\usepackage{pgfplots}
\pgfplotsset{compat=newest}
\usepgfplotslibrary{fillbetween}
\usetikzlibrary{arrows}
\usetikzlibrary{calc}
\def\centerarc[#1](#2)(#3:#4:#5)
{ \draw[#1] ($(#2)+({#5*cos(#3)},{#5*sin(#3)})$) arc (#3:#4:#5); }

\pgfdeclareplotmark{rottriangle}{%
  \pgfpathmoveto{\pgfqpoint{0pt}{-\pgfplotmarksize}}%
  \pgfpathlineto{\pgfqpointpolar{30}{\pgfplotmarksize}}%
  \pgfpathlineto{\pgfqpointpolar{150}{\pgfplotmarksize}}%
  \pgfpathclose
  \pgfusepathqstroke
}%
\pgfdeclareplotmark{rottriangle*}{%
  \pgfpathmoveto{\pgfqpoint{0pt}{-\pgfplotmarksize}}%
  \pgfpathlineto{\pgfqpointpolar{30}{\pgfplotmarksize}}%
  \pgfpathlineto{\pgfqpointpolar{150}{\pgfplotmarksize}}%
  \pgfpathclose
  \pgfusepathqfillstroke
}

\pgfplotsset{select coords between index/.style 2 args={
    x filter/.code={
        \ifnum\coordindex<#1\fi
        \ifnum\coordindex>#2\fi
    }
}}

\tikzset{
        hatch distance/.store in=\hatchdistance,
        hatch distance=10pt,
        hatch thickness/.store in=\hatchthickness,
        hatch thickness=2pt
    }
    \makeatletter
    \pgfdeclarepatternformonly[\hatchdistance,\hatchthickness]{flexible hatch}
    {\pgfqpoint{0pt}{0pt}}
    {\pgfqpoint{\hatchdistance}{\hatchdistance}}
    {\pgfpoint{\hatchdistance-1pt}{\hatchdistance-1pt}}%
    {
        \pgfsetcolor{\tikz@pattern@color}
        \pgfsetlinewidth{\hatchthickness}
        \pgfpathmoveto{\pgfqpoint{0pt}{0pt}}
        \pgfpathlineto{\pgfqpoint{\hatchdistance}{\hatchdistance}}
        \pgfusepath{stroke}
    }
    \makeatother

\usetikzlibrary{decorations.markings}
\tikzset{->-/.style={decoration={
  markings,
  mark=at position #1 with {\arrow{>}}},postaction={decorate}}
}
  \tikzset{-<-/.style={decoration={
  markings,
  mark=at position #1 with {\arrow{<}}},postaction={decorate}}
}

\pgfplotsset{
  colormap={mygreys}{rgb=(0,0,1) color=(black) rgb255=(238,140,238)},
}

\usepackage{glossaries}
\newacronym{dem}{DEM}{discrete element methods}
\newacronym{ps}{PS}{polystyrene}
\newacronym{ct}{CT}{computed tomography}
\newacronym{zarm}{ZARM}{Center of Applied Space Technology and Microgravity}
\newacronym{paa}{PAA}{poly acrylic acid}
\newacronym[plural=YSFs, firstplural=yield-stress fluids (YSFs)]{ysf}{YSF}{yield-stress fluid}
\newacronym{iss}{ISS}{International Space Station}

\begin{document}

\title{Spreading droplets of yield-stress fluids with and without gravity}

\def\dlr{\affiliation{%
Institute for Frontier Materials on Earth and in Space, German Aerospace Center (DLR), 51170 K\"oln, Germany}}
\def\hhu{\affiliation{%
Department of Physics, Heinrich-Heine Universit\"at D\"usseldorf, Universit\"atsstra{\ss}e 1, 40225 D\"usseldorf, Germany}}
\def\superaero{\affiliation{%
Institut Sup\'{e}rieur de l'A\'{e}ronautique et de l'Espace (ISAE-SUPAERO), Universit\'{e} de Toulouse, Toulouse, France}}
\def\uva{\affiliation{%
Van der Waals-Zeeman Institute, University of Amsterdam, Science Park 904, 1098 XH Amsterdam, The Netherlands}}
\author{Linnea Heitmeier}
\email{linnea.heitmeier@dlr.de}\dlr\hhu
\author{Olfa D'Angelo}\superaero\dlr
\author{Maziyar Jalaal}\uva
\author{Thomas Voigtmann}\dlr\hhu

\date{\today}

\begin{abstract}
We investigate the effect of gravity on the spreading of droplets of yield stress fluids, by performing 
both microgravity experiments (in a drop tower) and experiments under terrestrial gravity.
We investigate the dependence of the final droplet shape on yield stress and gravity.
Droplets are deposited on a thin film of the same material, allowing to directly test 
scaling laws derived from the thin-film equation for viscoplastic fluids.
Microgravity conditions allow to vary independently the two relevant
dimensionless numbers, the Bond number $\mathcal B$ and the plastocapillary
number $\mathcal J$, and thus to disentangle the influence of surface tension
from that of the yield stress on the droplet shapes.
Simulations using a visco-elastic model with shear thinning complement
the experiments and show good agreement regarding the droplet shapes.
Possible deviations arising in the regime of non-negligible elastic
effects and large plastocapillary numbers (large yield stress) are discussed.
\end{abstract}

\maketitle

\section{Introduction}

\Glspl{ysf} are characterized by their dual rheological response:
they flow like (complex) fluids only when the applied stress exceeds a
threshold (the yield stress). Below this threshold, their deformation
behavior becomes solid-like \cite{Balmforth.2014,bonn2017yield}.
This invites a wide range of applications:
many commonly used materials such as pastes, gels, emulsions, foams, and a variety of biological and industrial suspensions 
exhibit a yield stress \cite{Coussot}.
As a result, the spreading of \glspl{ysf} (and in general non-Newtonian fluids)
is central to many processes in sectors including food processing, cosmetics,
pharmaceuticals, construction, and energy. 
For example, 
the presence of a yield stress in 3D-printing materials ensures that extruded filaments or droplets retain their shape after deposition, enabling the controlled, layer-by-layer fabrication of complex structures \cite{placone2018recent, kyle2017printability, jiang2020extrusion, buswell20183d,van2023viscoplastic}.

Despite this practical importance, some fundamental aspects of the spreading
of \glspl{ysf} have only been established recently
\cite{cormier2012beyond,agrawal2019analysis,jalaal2021spreading,francca2024elasto}.
A defining characteristic of \glspl{ysf} is that their droplets reach a finite size even on fully wetting substrates. Unlike Newtonian fluids, which would theoretically spread indefinitely into an infinitesimally thin film, \glspl{ysf} balance hydrostatic pressure and surface tension against the yield stress, resulting in droplets of finite radius. 
This principle underpins direct-ink writing technologies, and notably bioprinting \cite{Paxton.2017,Townsend.2019,Milazzo.2023} both on ground and in space \cite{Ombergen.2023}.
We are therefore interested in how the asymptotic droplet size -- achieved at long times, $t\to\infty$
 after extrusion -- depends on parameters such as the deposited fluid volume, material properties, and gravity.

Sicne the formation and spreading of \gls{ysf} droplets is mainly governed by three factors -- yield stress, surface tension, and hydrostatic pressure --
(at least) two dimensionless groups are needed to characterize
their flow properties \cite{jalaal2021spreading}.
The Bond number, $\mathcal B$,
quantifies the importance of gravity over
surface tension, and is well known from the spreading of Newtonian fluids.
The existence of a yield stress implies another relevant dimensionless quantity,
the plastocapillary number, $\mathcal J$, to quantify the importance
of yield-stress effects over surface-tension ones.
For a droplet of typical length-scale $\mathcal L$ 
(to be defined more precisely further below)
and fluid mass density $\rho$, one defines
\begin{align}\label{eq:numbers}
  \mathcal B &= \frac{\rho g \mathcal L^2}{\hat\sigma}\,, &
  \mathcal J &= \frac{\tau_0 \mathcal L}{\hat\sigma}\,.
\end{align}
Here, $\tau_0$ is the yield stress, $\hat\sigma$ the surface tension,
and $g$ the gravitational acceleration. 
Specifically, scaling laws were derived for the standard viscoplastic Bingham model \cite{jalaal2021spreading} that predict the asymptotic droplet radius $R$ as a function of the dimensionless numbers. A particularly intriguing aspect of these laws is the emergence of distinct scaling exponents when $\mathcal{B}=0$, highlighting this regime as one where further experiental studies are desired.

To vary the dimensionless numbers $\mathcal B$ and $\mathcal J$,
the length scale $\mathcal L$ is arguably the parameter that is most easily
changed in Eqs.~\eqref{eq:numbers}.
However, this does not allow to change both numbers independently.
Especially realizing the limit $\mathcal B\to0$ \textit{via} $\mathcal L\to0$,
\textit{i.e.} in infinitely small droplets, implies both $\mathcal J\to0$
and obvious experimental limitations.
To access a broader range of the parameter space and vary $\mathcal B$ and
$\mathcal J$ independently, we opt to vary the effective gravitational
acceleration, $g$. 
This allows us to probe the limit $\mathcal B\to0$
while keeping $\mathcal J$ finite.

The study of droplets in the regime $g\to0$ has hitherto mostly focused on Newtonian fluids. 
Surface tension measurements on droplets under such effective microgravity conditions have a long-established history \cite{Passerone.2011}.
Sessile droplets have also been studied at $g\to0$ to asses variations
in contact angles and wettability \cite{Ababneh.2006,Diana.2012,baldygin2023effect,Mielniczuk2018}. 
The coalescence of sessile water drops has been
studied on the \gls{iss},
investigating the contact line
motion at small Bond numbers, 
while taking advantage of
the enhanced optical resolution brought
by large droplets \cite{McCraney.2022b,mccraney2022coalescence}.

Non-Newtonian droplet studies in the limit $g\to0$ have been proposed as a promising way to investigate fluid parameters, 
such as surface tension and viscosity, by means of levitation and droplet oscillations \cite{tamim2021oscillations}. 
However, to date, nearly all studies have been conducted under Earth gravity, 
primarily focusing on the spreading of sessile and impacting droplets \cite{saidi2010influence,gorin2022universal,Biroun.2023,francca2024elasto,mobaseri2025maximum}. 
More applied studies include experimental setups mimicking inkjet printing \cite{jung2013role} 
or Leidenfrost droplets \cite{bertola2009experimental}.

We present results from microgravity experiments; by this, we denote
experiments performed in a freely falling setup where gravitational acceleration effectively vanishes. 
In this regime, we test the scaling laws for the final droplet radius that were predicted by \citet{jalaal2021spreading}.
While previous experimental and simulation studies have explored these laws \cite{francca2024elasto,van2023viscoplastic}, the ability to independently vary both $\mathcal B$ and $\mathcal J$ in our microgravity experiments enables a far more stringent test of the theoretical predictions.
Our findings show agreement with the theoretical power laws, including their predicted asymptotic prefactors. Some systematic deviations appear that we discuss in the context of non-vanishing elastic effects.
Complementary simulations show that a shear-thinning visco-elastic fluid is a good model system for the observed droplet shapes.

\section{Methods}

\subsection{Experiments}

\begin{figure}
\centering
\includegraphics{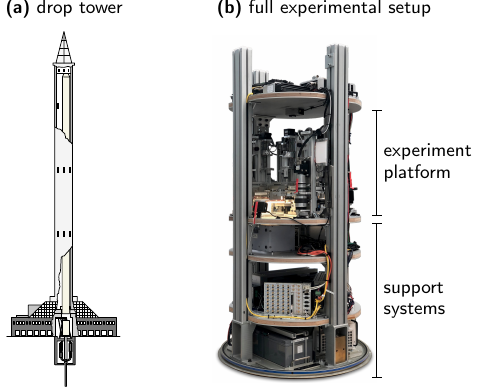}
\caption{\label{fig:setuptower} Platform used to conduct microgravity experiments.
(a)~Schematic of the \acrshort{zarm} drop tower, including catapult system below. 
(b)~Full experimental capsule, including experimental platform and support systems. 
}
\end{figure}

We conducted microgravity experiments ($g\to g_\mu\approx0$) complemented by
groud-based experiments (in nominal Earth gravitational acceleration,
$g_E=\SI{9.8}{\m\per\s}$).
Experimental campaigns were performed at the
\gls{zarm} drop tower in Bremen, Germany \cite{dittus1991drop, von2006new, dittus1990vacuum} (see Fig.~\ref{fig:setuptower}). 
Taking advantage of its catapult system, the \gls{zarm} drop tower provides a total of \SI{9.3}{\s} of microgravity ($g_\mu < 10^{-4}\,\si{\m\per\s}$). 
Ground reference
experiments were conducted in the same setup, using the same sample batches
as the corresponding microgravity experiment, and usually on the same day,
a few hours prior.
A detailed description of the hardware is provided elsewhere \cite{DAngelo2022}.
Below, we summarize the main steps of the experimental procedure.

\begin{figure*}
\centering
\includegraphics{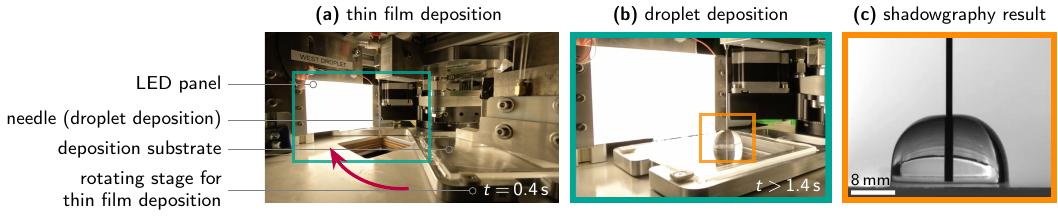}
\caption{\label{fig:setup}
Experimental setup and procedure for one droplet.
(a)~The thin film is deposited by rotating the substrate underneath a disposal blade (red arrow).
(b)~The droplet is deposited though a needle.
(c)~Exemplary shadowgraphy result.
}
\end{figure*}

The experiment consists of the deposition of a droplet of complex fluid on a
pre-wetted glass substrate covered by a thin film of the same fluid (see Fig.~\ref{fig:setup}).
The initial wetting of the substrate is done by rotating a rotary stage onto which the substrate is mounted, underneath a custom-designed nozzle-and-blade system.
The latter is designed to cover the glass surface with a thin fluid film.
Importantly, for the microgravity experiments, the thin film and droplet are both deposited during the microgravity phase \cite{DAngelo2022}.
Subsequently, a droplet is extruded through a nozzle placed in the center
of the droplet (inner diameter \SI{0.8}{\mm}, placed at \SI{0.8}{\mm} above the substrate). The glass surfaces are chemically treated to avoid slip~\cite{jalaal2015slip}.
The droplet volume, $V$, is controlled by a linear motor driving the extrusion syringe. 
The volumes deposited are $V=\SI{0.5}{\milli\liter}$, $\SI{1.8}{\milli\liter}$,
$\SI{4.2}{\milli\liter}$, and $\SI{5.6}{\milli\liter}$.
Different extrusion speeds were used, but found to have no
influence on the asymptotic droplet shapes discussed here.
For later reference, 
unless otherwise specified,
we give the corresponding length scale $\mathcal L$ as the diameter of a
sphere with the same volume, \textit{i.e.}, $\mathcal L=(6V/\pi)^{1/3}$.
The corresponding values are $\mathcal L=\qtylist{1.0;1.5;2.0;2.2}{\centi\meter}$.
(We omit the unit when quoting $\mathcal L$ values where convenient in
the following.)

Pre-wetting takes around
\SI{1.4}{\s}, and we find that the droplets attain stationary final shapes
well within the remaining microgravity time for these parameters.
The video recordings from all experiments are available online
\cite{zenodo1,*zenodo2}.

The droplets' spread is measured using 
shadowgraphy, a straightforward method for 
extracting the temporal evolution of axisymmetric droplet shapes. 
For this, the droplet is illuminated from one side using a custom-made LED panel, and observed from the opposite side using a high-speed camera.
The resulting images are subject to a standardized image-analysis
technique described in Appendix~\ref{apdx:image_analysis}.
We specifically extract the final droplet radius, $R$, and final height, $h$,
from these images. For the radius, two values are extracted from the left-
and right-hand part of the images, and both values are shown to indicate
the precision of the measurement.

\begin{table}
    \centering
    \small 
     \begin{tabular}{c c c} 
    \toprule
	{\quad\textbf{label}\quad} & {\quad\textbf{composition}\quad}  & {\quad\textbf{yield stress}\quad} \\
    \midrule
	c1 & 0.3~wt\%	&	\SI{6.5}{\pascal}		\\	
	c2 & 0.35~wt\%	&	\SI{9}{\pascal}		\\	
	c3 & 0.4~wt\%	&	\SI{14}{\pascal}		\\	
	c4 & 0.45~wt\%	&	\SI{21}{\pascal}	\\	
	c5 & 0.5~wt\%	&	\SI{35}{\pascal}		\\
	c6 & 0.55~wt\%	&	\SI{55}{\pascal}	\\
    \bottomrule
    \end{tabular}
\caption{ \label{tab:parameters_carbopol}	 
Concentration and yield-stress values of the Carbopol aqueous solutions used
in the experiments, with the labels used in the text and figure captions
as reference.
}
\end{table}

The complex fluids used as experimental materials are
Carbopol aqueous solutions, widely studied as model viscoplastic fluids \cite{barry1979rheological, curran2002properties, di2015characterization, bonn2017yield, frigaard2019simple, jalaal2019viscoplastic, jalaal2021spreading, Martouzet2021}.
Six different concentrations (labels c1 to c6) were used, ranging from
\SI{0.3}{\wtpercent} to \SI{0.55}{\wtpercent}, corresponding to
yield stresses, $\tau_0$,
between \SI{6.5}{\pascal} and \SI{55}{\pascal} (see Table~\ref{tab:parameters_carbopol} for the precise values).
Experiments cover a large (but not complete) set of combinations of droplet sizes and concentrations.
All solutions were produced by dilution from a \SI{1}{\wtpercent} stock
solution. The latter was prepared by dispersing \gls{paa} powder (Sigma Aldrich)
in Milli-Q water using a four-blade marine impeller rotating at
\SIrange{1000}{1500}{\rpm} at room temperature, and thereafter neutralizing
the solution to pH~7 with triethanolamine (Sigma Aldrich).
The yield stress values of the solutions are 
obtained using an Anton Paar rheometer (MCR-502), measuring flow curves
by increasing the shear rate
from $\SI{0.01}{\per\second}$ to  $\SI{1000}{\per\second} $ and fitting 
a Herschel-Bulkley model through the flow-curve data.

Note that from Eq.~\eqref{eq:numbers} one infers the droplet sizes for
which gravitational effects become relevant. This is usually expressed
through the capillary length
$\ell_\kappa=\sqrt{\hat\sigma/\rho g}$, such that $\mathcal B
=(\mathcal L/\ell_\kappa)^2\gg1$ if $\mathcal L\gg\ell_\kappa$.
For a typical liquid density of $\rho\approx\SI{1e3}{\kilo\gram\per\cubic\meter}$,
and with a the surface tension of water,
$\hat\sigma=\SI{0.072}{\newton\meter}$, as a reference,
we obtain $\ell_\kappa\approx\SI{2.7}{\milli\meter}$. All
our experimental data is in the regime $\mathcal L\gg\ell_\kappa$.

\subsection{Simulations}

The experiments are complemented by fluid dynamics simulations, 
carried out using the open-source tool \textsc{basilisk c}, a \textsc{C}-like
programming language adapted specifically to the task \cite{basilisk, popinet2009accurate}.

In \textsc{basilisk}, the volume of fluid technique is used to track the interface between the droplet fluid and a surrounding low-density, low-viscosity fluid (representing ambient air). 
A color function $f$ is used (where $f=1$ in the liquid and $f=0$ in the gas), 
which satisfies the scalar-advection equation.
One solves the governing equations using a one-fluid approximation \cite{prosperetti2009computational, tryggvason2011direct}, using
adaptive mesh refinement based on wavelet estimated discretization errors. The solver has been extensively used for various Newtonian and non-Newtonian problems with deformable interfaces~\cite{balasubramanian2024bursting,sanjay2021bursting,zhang2022impact,francca2024elasto,mobaseri2025maximum,jalaal2021spreading}.

The governing equations are given by 
the mass-conservation equation for incompressible fluids,
$\nabla\cdot\vec v=0$,
and the momentum conservation equation,
\begin{equation}
    \rho \frac{D}{Dt} \vec{v} = - \nabla p + \nabla \cdot \bs\sigma + \vec f_\text{ext}\,,
\end{equation}
where $\vec f_\text{ext}$ is an external force density representing
gravitational force, if present.
For the stress tensor, $\bs\sigma$, we use the constitutive
equation of a shear-thinning viscoelastic fluid that is representative
of a glass-forming fluid with a dynamical yield stress that emerges at the
glass transition \cite{Fuchs.2003,Papenkort.2015}.

Specifically, we set $\bs\sigma=\eta_\text{N}\bs{\dot\gamma}+\bs\sigma_p$,
where $\eta_\text{N}$ is a Newtonian background viscosity, and
$\bs{\dot\gamma}=(\nabla\vec v)+(\nabla\vec v)^T$ the symmetrized shear-rate
tensor. The non-Newtonian contribution is modeled in a form known as
a White-Metzner model from rheology,
\begin{equation}\label{eq:polymeric-part-def}
\frac{1}{\tau(\|\bs\dot\gamma\|)} \bs\sigma_p + \overset{\triangledown}{\bs\sigma}_p = G_\infty \bs{\dot\gamma}\,.
\end{equation}
For fixed relaxation time $\tau=\tau_\text{eq}$, this is the standard upper-convected
Maxwell model (Oldryod-B model including the Newtonian background viscosity).
The symbol $\triangledown$ indicates the upper-convected time derivative,
defined by
\begin{equation}
    \overset{\triangledown}{ A} = \frac{d}{dt}A -(\nabla \vec{v})^T \cdot A - A \cdot (\nabla \vec{v})
\end{equation}
for a twice-contravariant tensor $A$ such as the stress tensor
\cite{oldroyd1950formulation}.

The Oldroyd-B model describes viscoelastic stress relaxation: for time scales
$t\ll\tau_\text{eq}$, the response is that of an elastic solid, while for $t\gg\tau_\text{eq}$,
viscous flow sets in, with a non-Newtonian contribution to the viscosity
$\eta_p=G_\infty\tau_\text{eq}$.

To include the effect of shear-thinning, we introduce a shear-rate dependent relaxation rate: we set
\begin{equation}\label{eq:genmax}
\tau^{-1}=\tau_{eq}^{-1}+\frac{\|\bs{\dot\gamma}\|}{\gamma_c}
\end{equation}
in
Eq.~\eqref{eq:polymeric-part-def}. This form is inspired by microscopic
approaches to the rheology of glass-forming colloidal suspensions and
has been referred to as a ``nonlinear Maxwell model''
\cite{Fuchs.2003,Papenkort.2015}.
In the limit $\tau_\text{eq}\to\infty$, this model describes a
\gls{ysf}, with a yield stress $\tau_0=G_\infty\gamma_c$. For finite
$\tau_\text{eq}$, the model describes a quasi-\gls{ysf}, regularizing the
infinite zero-shear viscosity by a finite one. For the numerical simulations
considered here, we set $\tau_\text{eq}=\SI{50}{\s}$,
large enough to consider our simulations to be in the \gls{ysf} regime.
Asymptotic droplet shapes are measured in a time window where the initial
relaxation of the fluid has passed, but on a time scale $t<\tau_\text{eq}$
before viscous relaxation sets in. (See below for specific examples.)

The simulations consider droplet spreading in cylindrical symmetry,
assuming the $z$-axis to be the symmetry axis (and transforming the
Navier-Stokes equations into the corresponding coordinate system).
Cross-sections are thus obtained in the $(x,z)$-axis, and gravity,
if present, acts in the negative $z$-direction.
The initial conditions are that of a rotation
ellipsoid, given by
\begin{equation}
    x^2 + (\zeta z)^2 = R_0^2\,,
\end{equation}
with $\zeta=3$ to allow for the relaxation of prolate droplets to flatter
shapes.
The conversion between experimental values $\mathcal{L}$ and simulated radius $R_0$ is thus $\mathcal{L} = R_0 \cdot \left(\nicefrac{4}{\zeta}\right)^{1/3}$.
Further, a pre-wetted film of thickness $h_\infty=\SI{1}{\milli\meter}$
is added to the simulations for numerical stability.

As further simulation details,
the  fraction of densities is fixed as $\rho_\text{air}/\rho_\text{fluid}=0.001$,
and the fraction of Newtonian viscosities $\eta_\text{air}/\eta_\text{fluid}=0.1$. A box of side length $L_x=\SI{0.1}{\meter}$ (large enough to prevent finite size effects), and
a resolution level of 8 for adaptive mesh refinement in \textsc{basilisk} were chosen for the simulations.
The value of the surface tension $\hat\sigma=\SI{0.07}{\newton\per\meter}$
mimics that of a typical non-surfactant aqueous suspension.
For the constitutive equation, Eq.~\eqref{eq:polymeric-part-def},
$\gamma_c=1/5$ is fixed; this sets a typical strain scale where plastic
yielding starts to dominate over elastic response in the model, and values
of that order are to be expected on the grounds of microscopic theory
\cite{Fuchs.2003}.
To adjust the yield stress, we vary the Maxwell shear modulus,
$G_\infty=\SI{32.5}{\pascal}$, $\SI{45}{\pascal}$, $\SI{70}{\pascal}$,
$\SI{105}{\pascal}$, $\SI{175}{\pascal}$, and $\SI{275}{\pascal}$,
such that the simulated fluids have the same yield stress values as
determined for the Carbopol solutions used in the experiment.
Note that in this model, elasticity and yield stress are linked; the
addition of elastic effects is an extension of the simpler Bingham
viscoplastic model used in the theory \cite{jalaal2021spreading}.
Our viscoelastic shear-thinning model is also distinct from
elasto-viscoplastic models that incorporate an ad-hoc yield stress
and that have recently been used to study droplet spreading
\cite{francca2024elasto}.

\section{Results}

\subsection{Droplet shapes}

\begin{figure}
\centering
\includegraphics{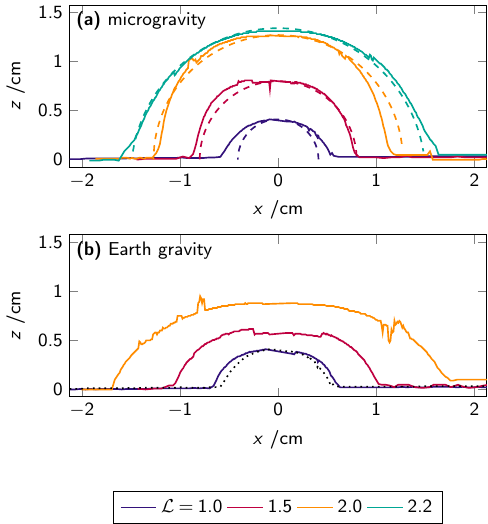}
\caption{\label{fig:shapes}
Experimental droplet shapes for different volumes
($\mathcal L=\SIlist{1;1.5;2;2.2}{\centi\meter}$ from bottom to top) for concentration c4.
Solid lines are reconstructed from image analysis, shown for
(a) microgravity experiments and (b) experiments in Earth gravity.
(Note that $\mathcal L=\SI{2.2}{\centi\meter}$ is missing in the latter.)
In (a), dashed lines correspond to half-spheres as a comparison.
In (b), a dotted line replicates the $\mathcal L=\SI{1}{\centi\meter}$ result from (a) to ease comparison.
}
\end{figure}

We begin by analyzing the droplet shapes obtained in the experiments.
Figure~\ref{fig:shapes} shows exemplary height functions obtained from the
image analysis, for a fixed concentration (fluid c4, \textit{cf.}\ Tab.~\ref{tab:parameters_carbopol}) and different droplet sizes,
for microgravity [Fig.~\ref{fig:shapes}(a)] and Earth gravity [Fig.~\ref{fig:shapes}(b)] experiments.
One can observe that the droplet shapes are clearly different depending on $g$.
For the concentration and range of parameters shown in the figure, the
droplets in microgravity are taller due to the
lack of hydrostatic pressure, and therefore consequently less wide at the same
volume than under Earth gravity.

All droplets considered here correspond to non-negligible Bond number.
For $\mathcal L=\SI{1}{\centi\meter}$ (smallest droplet), 
we get $\mathcal B\approx14$ in Earth gravity, 
or $\mathcal B_h=\mathcal B/4^{2/3}\approx5.4$ if we calculate the
Bond number from the expected height of the droplet. 
Comparing the droplet shapes in $g_\mu $ and $g_E$ (Fig.~\ref{fig:shapes}(a) to (b)),
we see that in this case, the droplet shapes are still little influenced by $g$.
For larger $\mathcal L$,
the gravity-induced flattening of the droplets becomes more and more apparent.
The largest droplets in microgravity (Fig.~\ref{fig:shapes}(a))
are reasonably close to half-spheres, indicated by dashed lines.
However, while this is true for this specific fluid,
it does not generalize to other concentrations --
we return below to a discussion of the concentration-dependence of droplet shapes.

\begin{figure}
\centering
\includegraphics{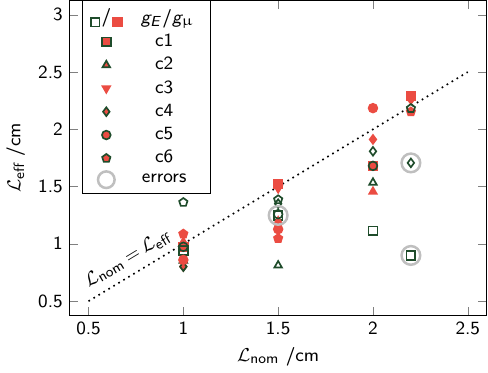}
\caption{\label{fig:shapeslopes}
Verification of the extruded volumes: equivalent sphere diameter $\mathcal L_\text{eff}=(6/\pi)V_\text{exp}$ from the numerically reconstructed droplet volumes,
as a function of their nominal size $\mathcal L$.
Full symbols correspond to droplets in microgravity
experiments, open symbols to those under Earth gravity; symbol
shapes indicate the various Carbopol concentrations as labeled. The
dotted line is the expected $\mathcal L_\text{eff}=\mathcal L$.
}
\end{figure}

While the droplet volume is in principle fixed by the automated deposition system,
we account for possible deviations in the finally extruded volume.
To this end, the droplet cross-sections extracted from the image analysis
are numerically integrated to obtain their volume. Assuming cylindrical
symmetry around the $z$ axis, we obtain $V_\text{exp}=2\pi\int_{x\gtrless0}x\,z(x)\,dx$.
In the experiments, we confirmed by shadowgraphy also from below for
some droplets their cylindrical symmetry (see Appendix~\ref{apdx:image_analysis} for details and images).
The integral is taken over either positive $x$ or negative $x$, resulting
in two estimates for $V_\text{exp}$. Note that the image analysis tends to
systematically underestimate $z(x)$: reflections from the light source
sometimes cause bright features inside the shadowgraphy projection of the
droplet. We thus obtain $V_\text{exp}$ as the larger of the two integrals.
This estimate of the extruded volume
defines the experimentally corrected droplet length scale,
 $\mathcal L_\text{eff}=(6/\pi)V_\text{exp}$. 

The resulting
$\mathcal L_\text{eff}$ are compared with the nominal $\mathcal L$
in Fig.~\ref{fig:shapeslopes}. We see that in particular in Earth gravity (open symbols),
the reconstructed droplet volumes sometimes show large deviations from the
expected ones, droplets often appearing smaller than expected. 
The microgravity data is much closer to the expected trend.

Discrepancies between $\mathcal{L}$ and $\mathcal{L}_\text{eff}$ arise from two main sources apart from poor reconstruction by image analysis. 
In the experiments, great care was taken to avoid air inclusions in the fluids, and the syringes were driven by precisely controlled linear motors. However, the tubing connecting the syringes and extrusion needles has finite stiffness, which can result in a quantity of extruded material that is lower than expected.
There may also be fluid loss into the pre-wetting film.
Both these sources of error are compatible with a systematic under-estimation
of $\mathcal L$.

\subsection{Scaling laws}

\begin{figure*}
    \centering
    \includegraphics[width=.8\linewidth]{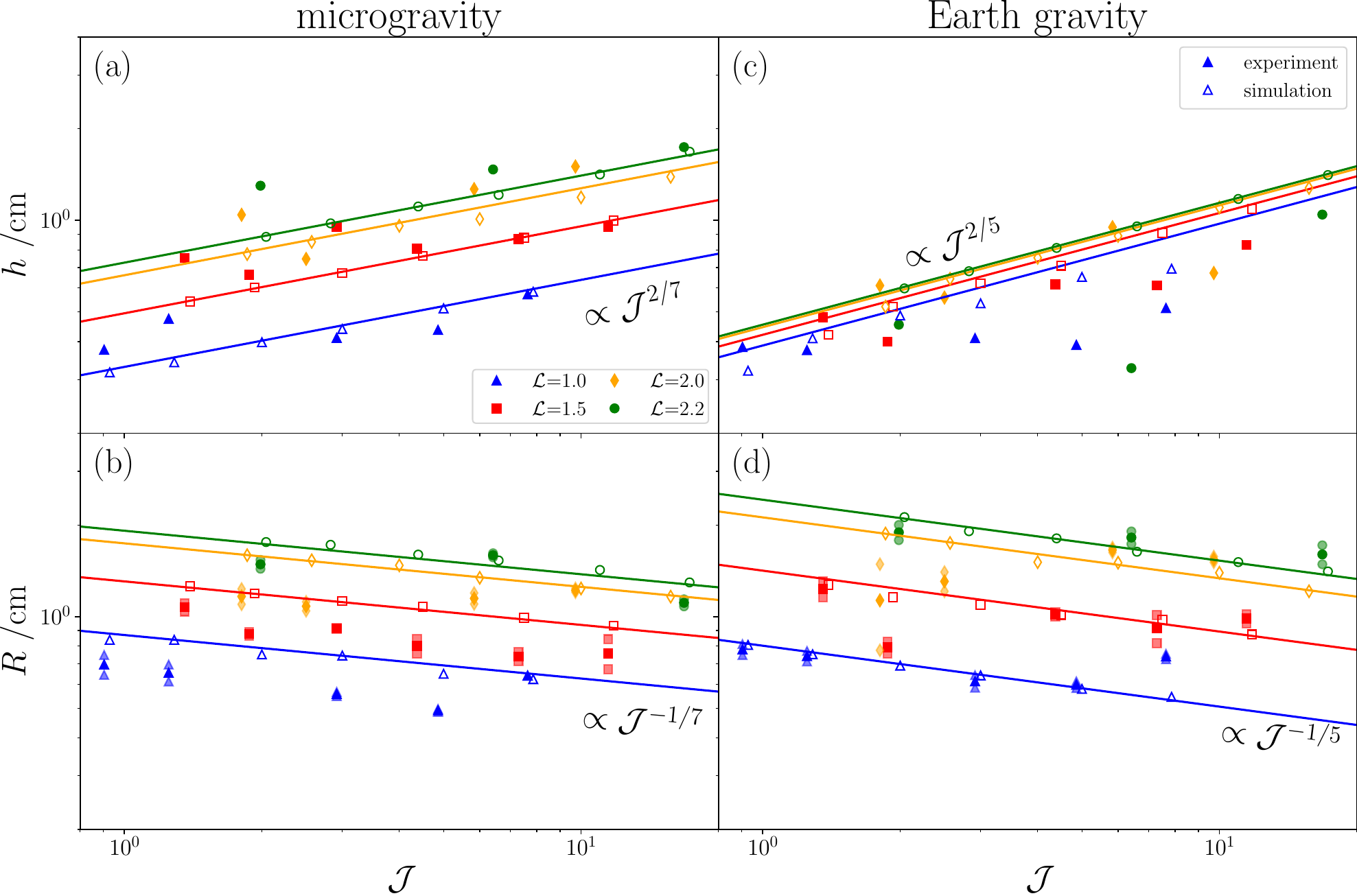}
    \caption{Test of the scaling laws of Ref.~\protect\cite{jalaal2021spreading}
      for final droplet sizes as a function
      of viscoplastic number $\mathcal J$, for different sizes $\mathcal L$.
      Filled symbols correspond to the experimental data ($\mathcal L=\qtyrange{1.0}{2.2}{\centi\meter}$ as labeled), open symbols to simulations for
      height (top row) and radius (bottom row); (a) and (b) refer to
      microgravity data, $\mathcal B\approx0$, while (c) and (d) refer to
      data for $\mathcal B\gg1$.
      Lines in all panels correspond to the expressions given in
      Eqs.~\protect\eqref{eq:scalingR} and \protect\eqref{eq:scalingh};
      for the height values, hemi-spherical droplets were assumed in
      evaluating a global prefactor.
      In (b) and (d) light-filled symbols indicate spearate estimates for
      the radii to indicate image-analysis error.
	\label{fig:scalings_exp}}
\end{figure*}

\citet{jalaal2021spreading}
derived scaling laws for
the dependence of the final droplet radius on the dimensionless numbers
$\mathcal B$ and $\mathcal J$. These scaling laws arise from an
asymptotic analysis of the thin-film equations, using the Bingham
model as a constitutive equation for the stress tensor.
One obtains for the final radius of the droplet
\begin{equation}\label{eq:scalingR}
  \frac{R}{\mathcal L}\simeq\begin{cases}
    \frac12\left(\frac{25}{32}\right)^{1/5}\mathcal B^{1/5}\mathcal J^{-1/5}\,,
    & \text{$\mathcal B\gg1$,}\\
    \beta\cdot\mathcal J^{-1/7}\,,
    & \text{$\mathcal B\to0$,}\end{cases}
\end{equation}
with $\beta\approx0.87$
(converting the results of Ref.~\cite{jalaal2021spreading} to our convention
that introduces a factor of $2$ in $\mathcal L$, as we define $\mathcal L$ as the diameter of the equivalent sphere, not its radius).
Accounting for conservation of fluid volume, in three dimensions, this
implies
\begin{equation}\label{eq:scalingh}
  \frac{h}{\mathcal L}\sim\begin{cases}
  4\alpha \left(\frac{32}{25}\right)^{2/5}
  \mathcal B^{-2/5}
  \mathcal J^{2/5}\,,
    & \text{$\mathcal B\gg1$,}\\
  \alpha\beta^{-2}
  \mathcal J^{2/7}\,, 
    & \text{$\mathcal B\to0$.}\end{cases}
\end{equation}
with a prefactor $\alpha$ that depends on the droplet shape and is
$\alpha=(3/2\pi)(\pi/6)=1/4$ for droplets of ellipsoidal shape with
a volume equal to a sphere of diameter $\mathcal L$.

These scaling laws are in principle free of fit parameters. However,
they require knowledge of the surface tension in the case $\mathcal B\to0$,
entering through the definition of the plasto-capillary number $\mathcal J$.
In the following we assume a fixed value of $\hat\sigma=\SI{72}{\milli\newton\meter}$,
assuming that the surface tension of the aqueous solutions is close enough
to that of pure water. We will comment on this assumption in
Sec.~\ref{sec:conclusion}.

Figure~\ref{fig:scalings_exp} shows the experimentally measured heights and
radii of the droplets in dependence of the non-dimensional number $\mathcal J$. 
Lines represent the scaling laws. For the case of the droplet radii,
the analytically calculated prefactors have been used, thus there are
no free fit parameters in this comparison.
We observe that all data are compatible with the two predicted power laws,
$R\sim\mathcal J^{-1/7}$ without gravitational acceleration,
and $R\sim\mathcal J^{-1/5}$
for the ground experiments. The height data follow the respective power laws,
however especially in the ground experiments, stronger deviations are seen,
including also the simulation data of $\mathcal L=\qty{1}{\centi\meter}$.
Note that the theoretical predictions for the height have to assume a certain
droplet shape, so that potentially some deviations can be attributed to this.
Specifically, we assume ellipsoidal shapes as discussed in conjunction
with Eq.~\eqref{eq:scalingh}, $\alpha=1/4$.
Possible sources of error coming from the experiment include the fact that
the droplets are extruded through a
needle that remains fixed in the center of the droplet. Hence capillary rise of the fluid around the needle cannot be completely avoided.
This influences both the droplet height and its
determination in image analysis. However, this would likely lead to
an over-estimation of the true height, whereas the deviations we observe
in Fig.~\ref{fig:scalings_exp}(c) are systematically towards lower values.

In the data from the microgravity experiments, we observe a systematic
deviation most notably in the droplet radii:
the experimentall determined
values fall below the analytical prediction. This effect had also been
seen in Ref.~\cite{jalaal2021spreading} where the limit $\mathcal B\to0$
could only be realized by letting $\mathcal L\to0$. Indeed, also in our
microgravity experiments, deviations appear to be most pronounced for the
smallest $\mathcal L$ [blue triangles in Fig.~\ref{fig:scalings_exp}(b)], while
the $\mathcal L=\qty{2.2}{\centi\meter}$ data appears to be somewhat closer
to the predcition (green circles).
However, the quality of our data does not allow to conclude a systematic
effect.
In the simulations (open symbols) the predicted analytical law
is quantiatively fulfilled for all $\mathcal L$; this would attribute
deviations for small $\mathcal L$ in the experiment to the details of
the extrusion mechanism and highlight the difficulty of analyzing
small droplets.

\begin{figure}
\centering
\includegraphics[width=.9\linewidth]{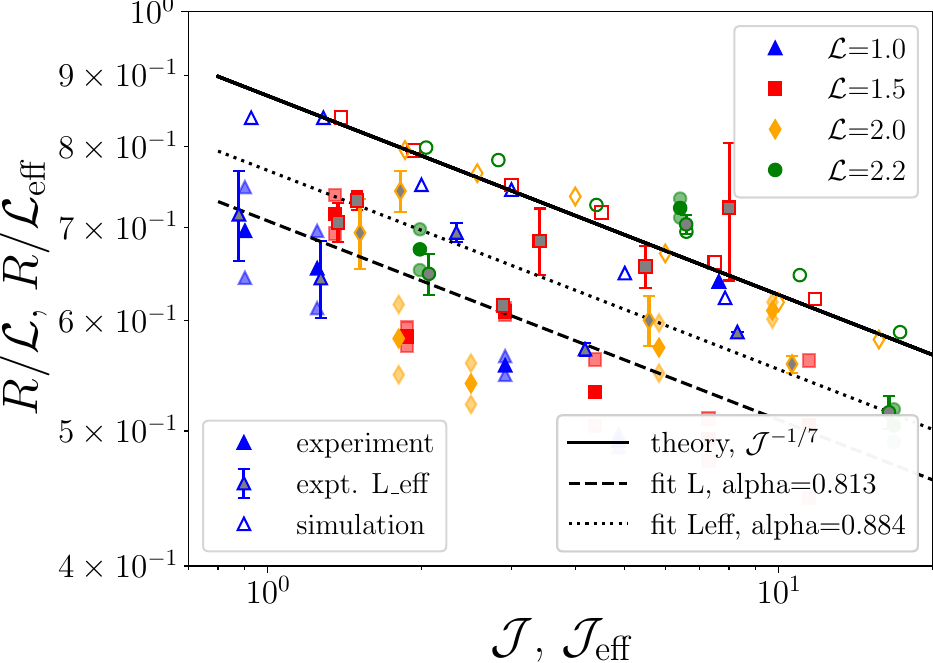}
\caption{\label{fig:mug_scaling}
  Scaling plot for the droplet radii in the regime $\mathcal B\to0$:
  normalized radii $R/\mathcal L$ are shown as a function of the
  plasto-capillary number $\mathcal J$.
  Filled and open symbols correspond data from microgravity experiments
  and simulations, respectively.
  A solid line represents the theoretical result, Eq.~\protect\eqref{eq:scalingR}.
  Symbols with error bars are experimental values corrected for the
  extruded volume (see text for details).
  A dashed (dotted) line represents the theory with an ad-hoc adjusted
  prefactor as noted in the legend, obtained from a fit of the
  experimental data without (with) this correction.
}
\end{figure}

Let us investigate the deviations in more detail.
Figure~\ref{fig:mug_scaling} repeats the radii obtained from the
microgravity experiments, but in the fully non-dimensionalized form
suggested by Eq.~\eqref{eq:scalingR}, \textit{i.e.}, $R/\mathcal L$
as a function of $\mathcal J$.
This confirms the data collapse predicted by the scaling law to
within \qty{20}{\percent} and highlights the systematic overestimation
of the radii by the theory.

In principle, a premature drying of the thin liquid film
on the substrate on which the droplet is spreading, can contribute to
a lesser spread than expected. However, we do not expect this to be a
systematic effect.
Incomplete extrusion as indicated by the fact that
$\mathcal L_\text{eff}<\mathcal L$ in the image analysis, could explain
a systematic deviation towards lower radii.
We thus show in Fig.~\ref{fig:mug_scaling} a comparison of the
data from the microgravity experiments to the scaling prediction,
where we also correct the nominal droplet sizes $\mathcal L$ to the
measured ones $\mathcal L_\text{eff}$ (symbols with error bars).
Note that this also shifts the dimensionless numbers to
corrected ones, $\mathcal J_\text{eff}$ and $\mathcal B_\text{eff}$.

From theoretical considerations of an elasto-viscoplastic model
\cite{francca2024elasto},
one concludes that elastic effects that are not considered in the
Bingham model behind Eq.~\eqref{eq:scalingR} should suppress spreading.
In the limit of weak elasticity in this model, the leading-order correction
can be captured by a reduction of the prefactor of the visco-plastic
power law. Such fits, introducing an \textit{ad hoc} prefactor
in Eq.~\eqref{eq:scalingR}, are shown in Fig.~\ref{fig:mug_scaling}
for the data both using the nominal and the experimental determined
value of the length scale. We observe that the deviation is on the
order of \qtyrange{10}{20}{\percent}.

Let us note that attributing these deviations to finite elasticity is
subtle: our simulations employ a model that also contains elasticity,
but in the form of a visco-elastic liquid-material model, with a dynamical
yield stress that emerges as the flow rate drops below the relaxation
rate of the material. The asymptotic droplet shapes obtained from the
model agree well with the prediction based on the
Bingham model, except for systematic deviations in the $\mathcal L=1$ case,
that can be attributed to the finite $h_\infty$;
for a more detailed explanation of the simulation analysis
we refer to Sec.~\ref{sec:simulations} below.
In contrast, in Ref.~\cite{francca2024elasto},
an elasto-viscoplastic solid-material model with a static yield stress.
For the model used in the simulations here, no in-depth theoretical
analysis is available yet.

\begin{figure}
\centering
\includegraphics[width=.9\linewidth]{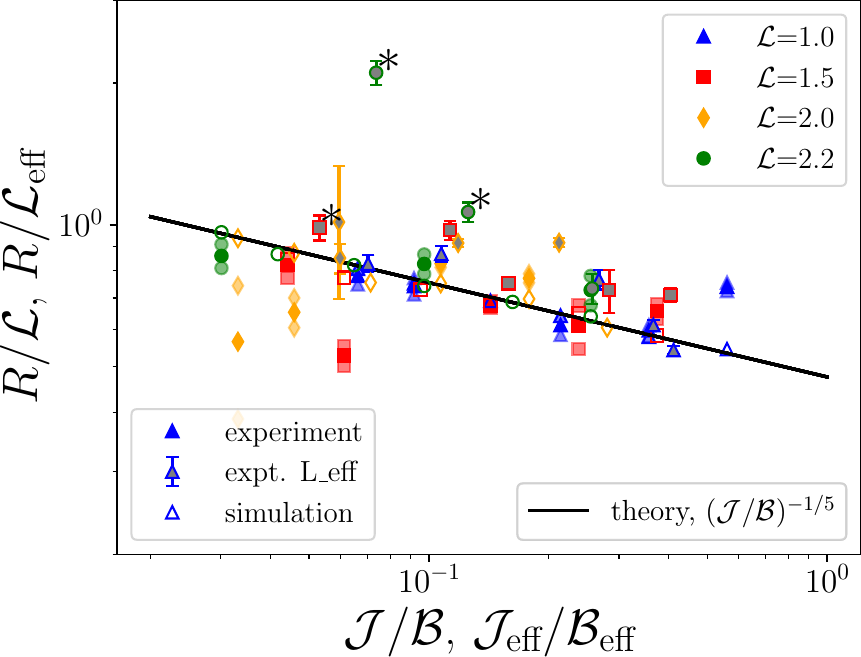}
\caption{\label{fig:g_scaling_JB}
  Scaling plot for the droplet radii in the regime $\mathcal B\neq0$:
  normalized radii $R/\mathcal L$ as a function of $\mathcal J/\mathcal B$.
  Filled and open symbols correspond to data from experiment and
  simulations, respectively. A solid line represents the
  theoretical result, Eq.~\eqref{eq:scalingR}.
  Symbols with error bars are experimental values corrected for
  $\mathcal L_\text{eff}$.
  Stars next to symbols indicate outliers as assessed by the quality
  of the image analysis.
}
\end{figure}

Also the gravity-based data collapses onto a master curve. According to
Eq.~\eqref{eq:scalingR}, the normalized radii $R/\mathcal L$ become a
unique function of $(\mathcal J/\mathcal B)$ for all droplet sizes.
This is verified with our data in Fig.~\ref{fig:g_scaling_JB} for which
we observe the predicted collapse to within experimental errors.
The possible correction to the nominal $\mathcal L$ coming from
our image analysis plays a sub-leading role here, as both data sets,
$R/\mathcal L$ as a function of $\mathcal J/\mathcal B$ (filled symbols
in Fig.~\ref{fig:g_scaling_JB}), and
$R/\mathcal L_\text{eff}$ as a function of $\mathcal J_\text{eff}/\mathcal B_\text{eff}$ (symbols with error bars),
do not show any systematic deviation from the prediction.

\begin{figure}
\centering
\includegraphics[width=.9\linewidth]{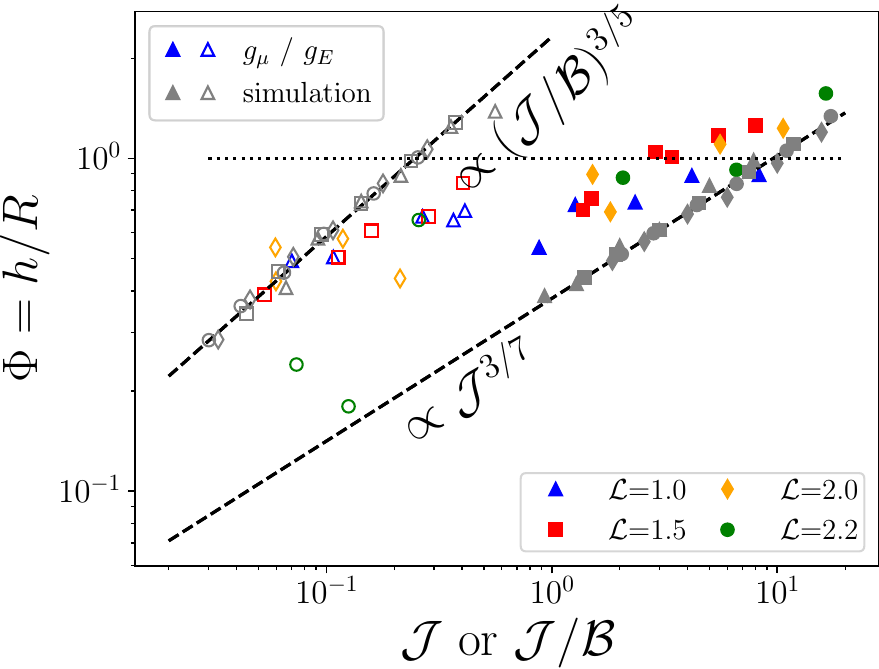}
\caption{\label{fig:aspectratio}Geometrical aspect ratio, $\Phi=h/R$, as a
function of the plastocapillary number (for the case in microgravity) and as a function of $\mathcal{J} / \mathcal B $ (for the case with gravity). Black dashed lines correspond to the theoretical predictions. Full symbols correspond to data from microgravity experiments, open symbols to those in Earth gravity. Grey symbols represent the simulation results.}
\end{figure}

The shape of the droplets, for fixed volume, depends on their yield stress
and on the presence of absence of gravity. We have extracted the
aspect ratio of the droplets, $\Phi=h/R$, from the experiments.
A value of $\Phi=1$ indicates a hemi-spherical droplet, while
for $\Phi<1$, droplet shapes are oblate, and $\Phi>1$ indicates
prolate droplets.

Combining the scaling laws,
Eqs.~\eqref{eq:scalingR} and \eqref{eq:scalingh}, one easily obtains
analytical predictions for the aspect ratio,
$\Phi\sim{\mathcal J}^{3/7}$ for the microgravity case $\mathcal B\to0$,
and $\Phi\sim{\mathcal J/B}^{3/5}$ in the presence of gravity, $\mathcal B\gg1$.
We show these predictions together with the aspect ratios from our
experiments in Fig.~\ref{fig:aspectratio}. For the analytical prediction,
we assumed ellipsoidal droplet shapes (setting $\alpha=1/4$) but used no
fitting parameter otherwise.
Both these asymptotes imply $\Phi\gg1$ for large enough $\mathcal J$.
For the microgravity conditions, our experimental values span the
regime of both $\Phi<1$ and $\Phi>1$.
In the ground-based experiments, we observe
$\Phi<1$ for all droplets,
\textit{i.e.}, they are all oblate
even for those combinations of $\tau_0$ and
$\mathcal L$ where the scaling laws suggest to reach the regime of prolate
droplets also under gravity.

Note that the numbers $\mathcal J$ and $\mathcal J/\mathcal B$ are
proportional to the yield stress $\tau_0$. Thus
from the droplets' aspect ratio shown in Fig.~\ref{fig:aspectratio},
we confirm previous results by \citet{german2009impact}:
a higher yield stress always increases droplets' aspect ratio.
This correlation can be attributed to the fact that a higher yield stress results in a larger immobile regions
inside the droplet, which can be lifted up while remaining immobile during deposition.
The final droplet shape hence becomes elongated upwards. 

\subsection{Comparison to Simulations}\label{sec:simulations}

\begin{figure}
\centering
\includegraphics[width=\linewidth]{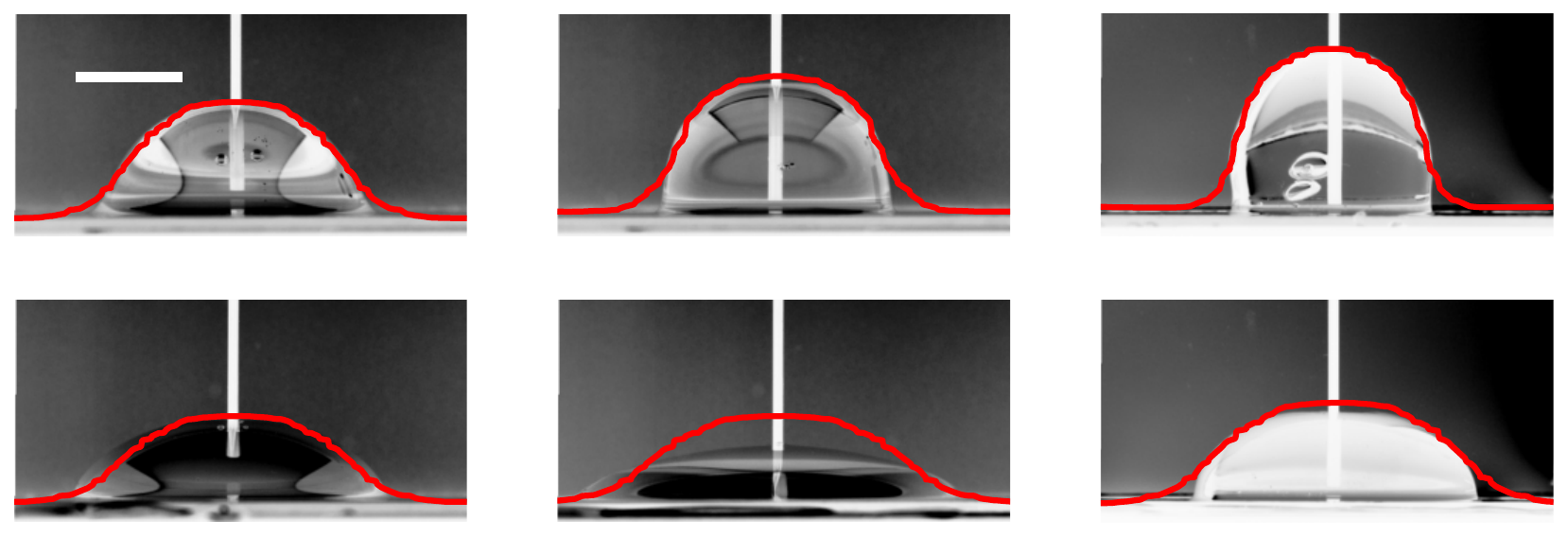}
\caption{\label{fig:ratio-illustration} Illustration of droplet shapes
  for droplets with size $\mathcal L=\SI{2.2}{\centi\meter}$: The top row
  shows images from microgravity experiments; the bottom row the
  corresponding results on ground. Different Carbopol concentrations
  are shown, with increasing yield stress from left to right (concentrations
  c1, c4, and c6).
  The white scale bar in the top left panel represents \qty{1}{\centi\meter}.
  Red lines indicate results from simulations with a shear-thinning
  Maxwell fluid, using various shear moduli $G_\infty$ to match the
  experimental yield-stress values.
  The large deviation in the center bottom panel is attributed to
  an experimental problem during droplet extrusion.
}
\end{figure}

Figure~\ref{fig:ratio-illustration} shows exemplary cases of droplets
for $\mathcal L=\SI{2.2}{\centi\meter}$ and different yield stresses
(Carbopol concentrations).
In the microgravity experiment one clearly sees the evolution towards more
prolate droplets with increasing yield stress.

Also shown in Fig.~\ref{fig:ratio-illustration} are simulation results
for asymptotic droplet shapes (red lines).
They confirm that our viscoelastic shear-thinning fluid model
is able to capture the specific droplet shapes well.
Deviations are seen towards to bottom. But note that the simulations use
a rather high background-fluid layer for numerical stability, which results
in droplet shapes that are much more ``rounded'' than the ones in experiment
towards the bottom, missing a clearly defined contact line.
In fact, in the experiment the apparent contact angles increase with
increasing yield stress \cite{DAngelo2022}, and they increase in the
microgravity experiment compared to the one in Earth gravity. For the
largest yield stress shown in Fig.~\ref{fig:ratio-illustration},
the apparent contact angle exceeds \ang{90}; an effect not observed in
the simulation.

\begin{figure}
    \centering
\includegraphics{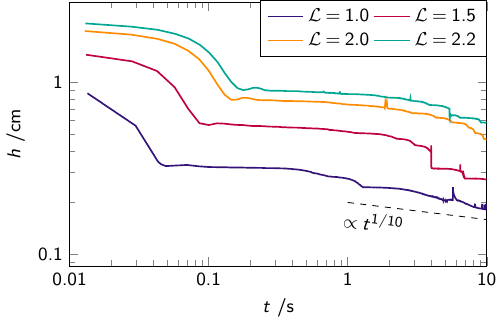}
    \caption{Height evolution as a function of time in the simulations;
    shown for a yield stress of $\tau_0=\qty{6.5}{\pascal}$ (c1), and
    different droplet sizes $\mathcal L$ as labeled.}
    \label{fig:heightdev}
\end{figure}

To illustrate how the asymptotic droplet shapes are extraced from the simulation,
we show in Fig.~\ref{fig:heightdev} exemplary results for the evolution
of the droplet height in the simulation. Results are shown for fixed
yield stress without gravity, and different
$\mathcal L$.
Typically for a visco-elastic fluid model, the height shows an initial
decrease at short times, driven by surface tension adjusting the shape
of the droplet from the arbitrarily chosen initial configuration.
For large droplets, this evolution hints at
some oscillations (around $t=\qty{0.2}{\s}$ in the figure for $\mathcal L=2.0$ and $\mathcal L=2.2$), indicative of elastic restoring forces.
After this initial decay, a plateau is observed in the height-\textit{versus}-time
plot. Recall that our model always allows flow on time scales $t\gg\tau$,
but acts like a yield-stress fluid for $t\ll\tau$. We thus extract
the asymptotic droplet shape that represents the balance of the yield stress
with the other forces (surface tension and gravity) in the plateau regime
that is (within the limits that we checked) independent of the precise initial
conditions and not yet affected by the ultimate liquid-like flow of the model.




\begin{figure}
    \centering
\includegraphics{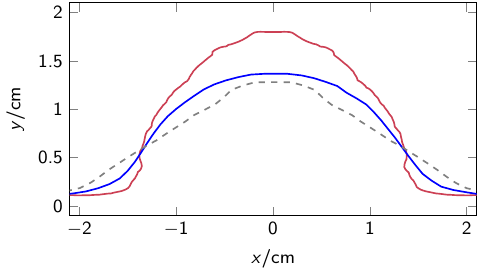}
    \caption{Shape of a droplet for $G_\infty = \SI{32.5}{\pascal}$ and $R=0.02 m$. Red: simulation results for the case without shear-thinning, blue: simulation results for the case with shear-thinning ($\gamma_c=1/5$). The case without shear thinning, but effective reduced shear modulus $G_\infty^\text{eff}=G_\infty\gamma_c$ is shown as a dashed line for comparison.}
    \label{fig:shape-without-yield}
\end{figure}

The simulation allows to address the role of the fluid rheology in determining
the droplet shapes. Recall that the scaling laws were obtained for a model
viscoplastic fluid, where a yield stress is incorporated in an ad-hoc manner.
The good agreement with our simulations of a refined model of a visco-elastic
shear-thinning fluid is encouraging: in the asymptotic long-time regime
that we analyze, only the value of the (dynamic) yield stress $\tau_0$ enters.

On purely dimensional grounds, also elastic effects might be considered,
\textit{i.e.}, the balance of elastic restoring forces (quantified by the
shear modulus $G_\infty$ of the viscoelastic fluid) compared to
surface-tension forces.
However, the droplet shapes observed in experiment are clearly determined
by the dynamic yield stress. To assert this, we compare in
Fig.~\ref{fig:shape-without-yield} droplet shapes obtained from the
simulation with and without shear thinning.
We observe that the purely viscoelastic droplet (red line in
Fig.~\ref{fig:shape-without-yield}) remains taller and less spread
compared to the one with shear thinning: shear thinning initially aids the flow
of the material in response to the surface tension forces. The forces
opposing the complete spread of the droplet are then given by the yield
stress, $\tau_0=G_\infty\gamma_c$, and these are not equivalent to a
simply reduced elastic force $G_\infty^\text{eff}$: as the comparison
with the model without shear thinning, but $G_\infty^\text{eff}
=G_\infty\gamma_c$ shows (dashed line), the droplet shapes are not
identical with and without shear thinning.

\section{Conclusion}\label{sec:conclusion}

We studied the spreading of droplets of Carbopol suspensions, a model
yield stress fluid, both with and without gravitational acceleration.
The droplets spread on a pre-wetted thin film, and eventually attain a finite shape, 
due to a balance between surface tension, their yield stress,
and hydrostatic pressure in the presence of gravity.
Conducting microgravity experiments at the \gls{zarm} drop tower, 
we were able to
separate the limit $\mathcal B\to0$ from the limit $\mathcal J\to0$
by decoupling $\mathcal{B}$ from the droplet size, $\mathcal L$.
We could hence verify scaling laws by \citet{jalaal2021spreading}
for both the gravity-dominated case (characterized by finite Bond number,
$\mathcal B\gg1$) and the surface-tension dominated case
($\mathcal B\to0$, at finite plastocapillary number $\mathcal J$).
We complemented our experiments by simulations using a model including
a dynamical yield stress together with visco-elasticity, and found that
the simulations obey the scaling predictions very well.

The scaling laws predict a power-law dependence of the final droplet size on
the plastocapillary number, $\mathcal J$. For the experimental
droplet radii we find good
agreement with the predictions, while the droplet heights show somewhat
larger deviations, especially in the gravity-dominated case.

Using a standard viscoplastic constitutive equation, \textit{viz.} the
Bingham model, \Citeauthor{jalaal2021spreading}
also derived the numerical values of the prefactors in the scaling laws.
Our experimental results confirm these prefactors, but show that
in order to observe them, the limit $\mathcal L\to0$ should be avoided
if $\mathcal B\to0$.
Systematic deviations could point to finite elasticity effects
as discussed more recently by \citet{francca2024elasto}.
There, an extension of the Bingham model to include both a static
yield stress and solid elasticity was studied, and corrections to the
scaling laws coming from finite Ohnesorg and Deborah number
would be compatible
with our experimental data.
Note however that our simulations show good agreement with the
theoretical predictions based on the Bingham model, even though our
simulation model includes visco-elastic effects. This is a suble point:
our model is ultimately a liquid-like model, with a regularized
\emph{dynamical} yield stress. The model of \Citeauthor{francca2024elasto}
in contrast has a \emph{static} yield stress and included elasticity
in a solid-like model. The connection between these different
constitutive modelling approaches will have to be investigated further.

A potential cause of deviations from the scaling laws could be hidden
in the role of the surface tension. In our analysis, we have assumed all
our Carbopol solutions to have the same surface tension as water, irrespective
of their yield stress. This is in line with recent experiments
\cite{Mohammadigoushki.2023},
and the interpretation that Carbopol as a non-surfactant polymer should
not drastically change the surface tension of the solvent.
Note however that the determination of surface tension in the presence
of a yield stress is intricate \cite{boujlel2013measuring, jorgensen2015yield},
as the \gls{ysf} approaches
solid-like behavior and one in principle needs to distinguish surface
tension and surface energy.

Making use of the scaling laws for the case $\mathcal B=0$ one could
in principle determine the surface tension from a measurement of the droplet
radii or heights,
\begin{equation}\label{eq:sigma_from_R}
  \hat\sigma = \left(\nicefrac{R}{\beta}\right)^7\mathcal L^{-6}\tau_0
\end{equation}
respectively
\begin{equation}\label{eq:sigma_from_h}
  \hat\sigma = \left(\nicefrac{1}{2\beta\sqrt{h}}\right)^7{\mathcal L}^{9/2}\tau_0\,.
\end{equation}
In principle, data from microgravity experiments would be uniquely suited
to determine $\hat\sigma$ this way, because the expected variation with
$\mathcal L$ can be tested for fixed $\tau_0$ and not leaving the
required $\mathcal B\to0$ regime. However, the high powers of $R$ and $h$
entering the equations cause large uncertainties. For our data, the
obtained values of $\hat\sigma$ are still compatible with the value
of the pure solvent.

\section*{Data Management}

The data associated to this study is available on Zenodo repositories 15806543 and 15806731 \cite{zenodo1,*zenodo2}.

\begin{acknowledgments}

We warmly acknowledge engineering support from the Technical Center from the University of Amsterdam, 
in particular Kasper van Nieuwland, Clint Ederveen Janssen Tjeerd G.L.C.\ Weijers and Daan Giesen,
as well as the entire \gls{zarm} support team. 

The authors gratefully acknowledge the scientific support and HPC resources
provided by the German Aerospace Center (DLR). The HPC system CARO is
partially funded by \enquote{Ministry of Science and Culture of Lower Saxony} and
\enquote{Federal Ministry for Economic Affairs and Climate Action}.

O.\ D'A.\ acknowledges financial support from the 
European Low Gravity Research Association (ELGRA), who initially 
supported this project through the 2021 ELGRA Research Prize, and
French National Centre for Space Studies (CNES) under the CNES fellowship 24-357.
\end{acknowledgments}

\bibliography{bib.bib}
\newpage

\section*{Appendix}


\appendix

\section{Data analysis of the experimental data}\label{apdx:image_analysis}

\begin{figure}
    \centering
    \subfigure[Step 1 \label{fig:Step1}]{\includegraphics[width = 0.3 \linewidth]{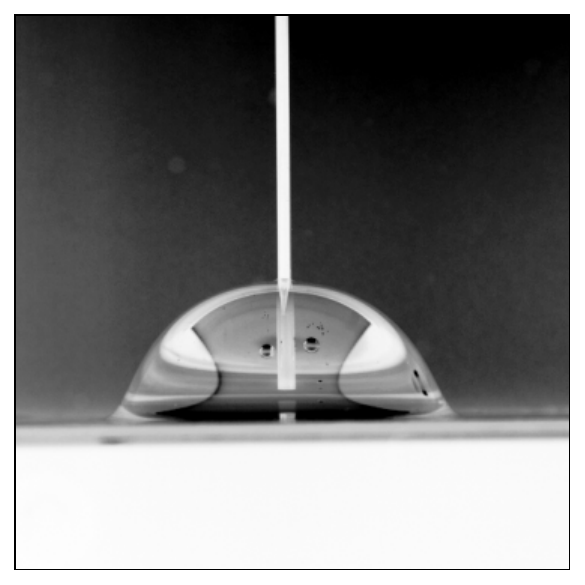}}
     \subfigure[Step 2 \label{fig:Step2}]{\includegraphics[width = 0.3 \linewidth]{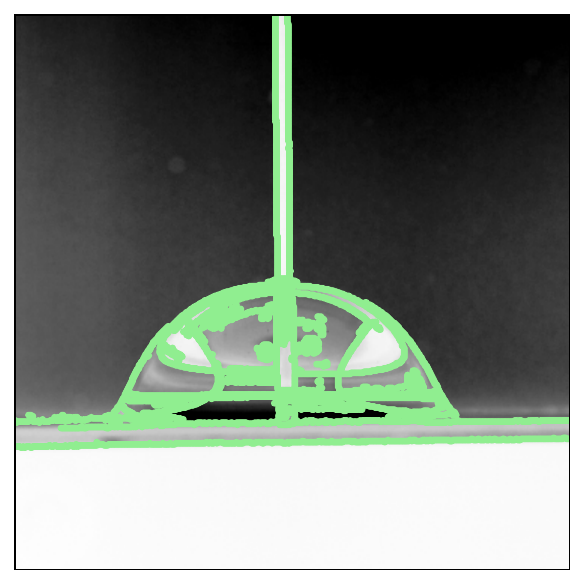}}
      \subfigure[Step 3 \label{fig:Step3}]{\includegraphics[width = 0.3 \linewidth]{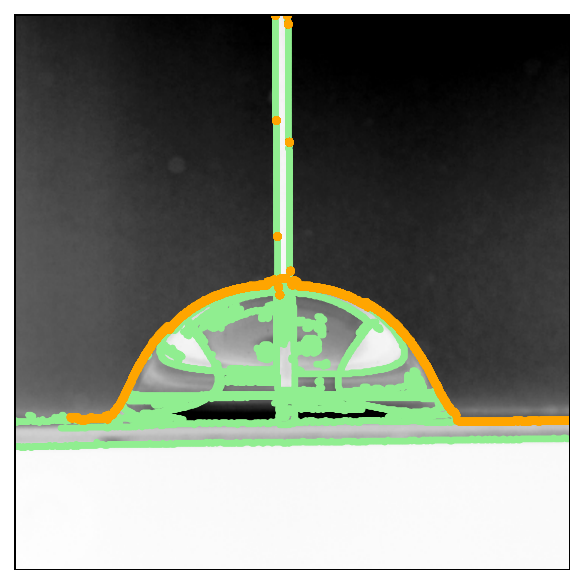}}
       \subfigure[Step 4 \label{fig:Step4}]{\includegraphics[width = 0.3 \linewidth]{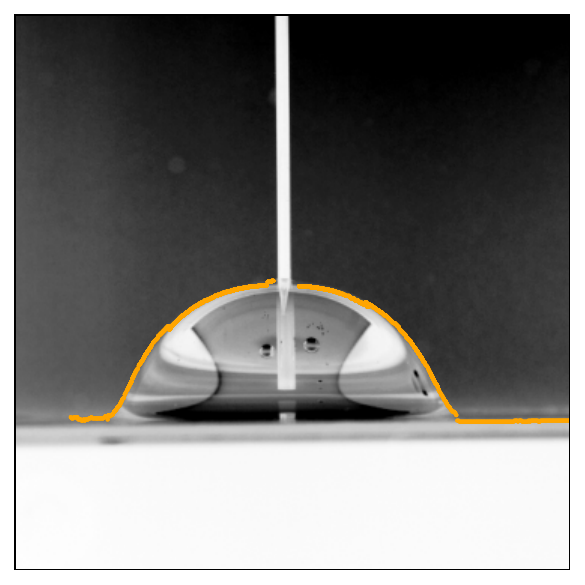}}
        \subfigure[Step 5 \label{fig:Step5}]{\includegraphics[width = 0.3 \linewidth]{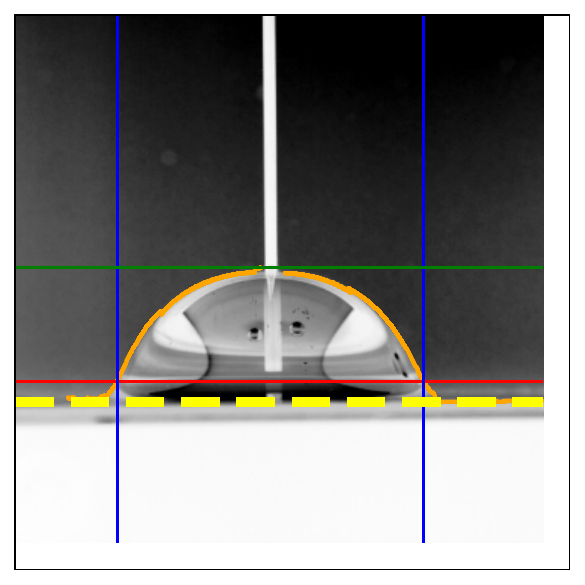}}
    \caption{Steps in the image analysis. See text for a deeper description.}
    \label{fig:image_analysis}
\end{figure}

We briefly summarize our image analysis procedure to obtain the droplet
shapes. We analyze the images from the experiment in 5 different steps, as sketched in Figure~\ref{fig:image_analysis}.

From the video recording of the experiment, we extract the raw image of the
droplet in the late stage of the experiment (just before impact in the microgravity case), as in Fig.~\ref{fig:Step1}.
We first identify all edges in the images with a canny-edge filter
(standard OpenCV implementation in python, \texttt{cv2.Canny}, using
thresholds of 48 \% and 52\% of the maximal image pixel.
This step is illustrated in Fig.~\ref{fig:Step2}.
To keep only the outer shape of the droplet, all edges detected inside the
droplet (typically from small air inclusions or reflections) are removed.
This is done by deleting, for every $x$-values, every $y$-value except the highest value.
The outer shape of the droplet thus found is marked yellow in Fig.~\ref{fig:Step3}.
As the deposition needle, which is part of the experimental images, should not be part of our analysis,  we then delete all values which have an $x$-value in the range of that of the needle. The resulting image can be seen in Fig.~\ref{fig:Step4}.

Finally, we perform the analysis of the extracted droplet shape. 
The bottom of the droplet (thin film of pre-wetting fluid), placed at height $h_0$, is found 
by taking the minimum value of the $y$-values of the droplet shape.
This is marked by the dashed yellow line in Fig.~\ref{fig:Step5}. 
The height of the droplet, $h$, can be extracted by taking the maximum value of the $y$-values of the droplet's shape. 
The height of the droplet is drawn as a green line in Fig.~\ref{fig:Step5}.
The radius is determined by defining a new height, $h_{\text{radius}}$, given by $h_0+ (h-h_0)\cdot 0.1$, which is drawn in red in Fig.~\ref{fig:Step5}. 
This rise of 10\% of the droplet's height from its bottom allows us to capture the droplets' radii without the wetting angle.
 We identify the $x$-values at which the shape of the droplet crosses this point to capture the droplet's extremities (blue lines in Fig.~\ref{fig:Step5}).

The center of the image is defined by the position of the needle.
To convert the values of the heights and radii obtained in image
processing, before each experiment calibration images with a $\SI{2}{\milli\meter}$ grid were taken; the resolution of the setup is typically $\SI{15}{\px\per\mm}$.

\begin{figure}[h!]
    \centering
\includegraphics{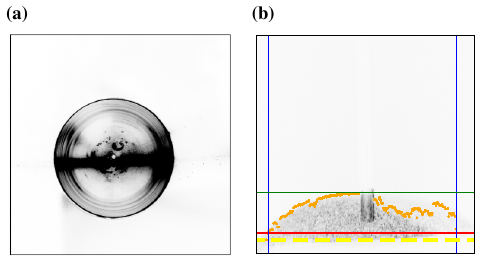}
    \caption{Left: Image from the bottom from the droplet of concentration c4 and $\mathcal{L}=2.2$. Right: Exemplary image, where the shape detection did not work perfectly.}
    \label{fig:image-analysis-risks}
\end{figure}

We verified the circularity of the droplets by pictures taken from below the glass substrate (example in Figure~\ref{fig:image-analysis-risks}(a)).

Another concern in the volume reconstruction by image analysis 
is the fact that light reflections sometimes affect the detection of the droplet's contour.
One such example is given in Figure~\ref{fig:image-analysis-risks}(b): on the right side of the image, the image analysis algorithm picks up a light reflection instead of the true droplet's shape. 
In such case, we used the well reconstructed half of the droplet (left half here) and determined its volume by mirroring it.

\end{document}